# Time in the theory of relativity: on natural clocks, proper time, the clock hypothesis, and all that

Mario Bacelar Valente

Abstract

When addressing the notion of proper time in the theory of relativity, it is usually taken for granted that the time read by an accelerated clock is given by the Minkowski proper time. However, there are authors like Harvey Brown that consider necessary an extra assumption to arrive at this result, the so-called clock hypothesis. In opposition to Brown, Richard TW Arthur takes the clock hypothesis to be already implicit in the theory. In this paper I will present a view different from these authors by taking into account Einstein's notion of natural clock and showing its relevance to the debate.

1 Introduction: the notion of natural clock

Up until the mid 20$^{th}$ century the metrological definition of second was made in terms of astronomical motions. First in terms of the Earth's rotation taken to be uniform (Barbour 2009, 2-3), i.e. the sidereal time; then in terms of the so-called ephemeris time, in which time was calculated, using Newton's theory, from the motion of the Moon (Jespersen and Fitz-Randolph 1999, 104-6). The measurements of temporal durations relied on direct astronomical observation or on instruments (clocks) calibrated to the motions in the 'heavens'. However soon after the adoption of a definition of second based on the ephemeris time, the improvements on atomic frequency standards led to a new definition of the second in terms of the resonance frequency of the cesium atom. Motions within the solar system were set aside as the ultimate motion from which all motions were compared and time was measured; a new time scale was put forward in terms of a new definition of second based on a specific number of atomic transitions of the cesium atom and a new instrument to measure these transition (i.e. to measure 'time') was developed: the atomic clock (Jespersen and Fitz-Randolph 1999, 53-61).
  In a way this change in the metrology of time was anticipated in developments in theoretical physics. In the late 19$^{th}$ and early 20$^{th}$ century several thinkers were involved with issues related to the so-called electrodynamics of moving bodies. In his criticism of Lorentz's electron theory and its extension to the case of matter in (inertial) motion, Poincaré noticed that what for Lorentz was a mathematical artifice – that of rewriting his equations for the case of moving bodies in terms of auxiliary variables one of which Lorentz had called local time –, could have a completely different physical interpretation. According to Poincaré the local time can be the time being measured by observers in motion with the material bodies in question that synchronize their clocks by exchanging light signals (Poincaré 1900, 272). This procedure for synchronizing clocks had been presented by Poincaré, for the particular case of clocks taken to be at rest, in an earlier work published in 1898. In this work Poincaré mentions that even the best clocks, by that time still mechanical clocks, had to be calibrated to the sidereal time (Poincaré 1898, 3). We can still read Poincaré's 1900 remarks in the light of his 1898 memoir, i.e., implicitly, the clocks are taken to have been calibrated to sidereal time. However, due to the issue of the setting of the initial phase of distant clocks (the synchronization of the clocks), Poincaré discusses the relation between the time intervals read by the clocks in an apparently autonomous way.

This trend is stronger in Einstein's 1905 paper on the electrodynamics of moving bodies, nowadays known as the paper in which made its appearance the theory of relativity.[1] From the start Einstein develops his views in terms of measuring rods and clocks belonging to different inertial reference frames in relative motion. In this paper Einstein considers clocks "of exactly the same constitution" (Einstein 1905, 142). This means that the clocks are taken to have the same rate from the start (see, e.g., Einstein 1916, 273). Several things must be mentioned at this point. There is a threefold aspect to the notion of clock as used by Einstein. On one part, the term 'clock' refers to a particular instrument, e.g. the balance wheel clock, i.e. to real clocks (Einstein 1905, 153).[2] But also, as Einstein's admits in later writings, 'clock' is employed as an independent concept in the development of the theory (Einstein 1921, 212-3). This connects with the third aspect I want to mention. As a dynamical system the clock must be taken to be in agreement with the predictions of the theory, i.e. even if the clock is inbuilt into the theory as an independent concept in the definition of the inertial reference frame, it must be consistent with the dynamics built on top of the kinematics defined in part in terms of the inertial reference frames. Ultimately, Einstein's view is that this situation is not desirable:

In the present stage of development of theoretical physics these concepts must still be employed as independent concepts; for we are still far from possessing such certain knowledge of the theoretical principles of atomic structure as to be able to construct solid bodies and clocks theoretically from elementary concepts. (Einstein 1921, 213)

One is struck that the theory … introduces two kinds of physical things, i.e. (1) measuring rods and clocks, (2) all other things … strictly speaking measuring rods and clocks would have to be represented as solutions of the basic equations …, nor as it were, as theoretically self-sufficient entities. (Einstein 1969, 59-61; see also Giovanelli 2014)[3]

---

[1] In this work it is adopted Fock (1959) terminology. Instead of adopting special and general relativity to name Einstein's two theories, one refers to the 1905 theory of relativity and the 1915 theory of gravitation. The discussion on natural clocks, proper time and the clock hypothesis is made only in reference to the theory of relativity. It is beyond the scope of this work to address these issues in the context of Einstein's theory of gravitation.

[2] As such, even if Einstein does not address this issue, we must take the clock as ultimately calibrated to sidereal time, unless there is an argument to do otherwise; Einstein himself makes explicit in later writings the notion of natural clock, which I see as anticipating in part the later adoption of the atomic time scale based on atomic clocks.

[3] Einstein early philosophy is usually considered as a form of empiricism, to some close to what later become known as operationalism, even if there is disagreement regarding Einstein's operationalist sympathies (see, e.g., Paty 1993, 349-365, Scheibe 2001, 72-73; Dieks 2010, 228-35). On first sight Einstein's view of rods and clocks as independent conceptual elements of the theory might seem to be an example of this philosophical position. That is not the case. To Einstein, ideally, a physical theory must conform to Poincaré's conventionalist thesis (see, e.g., Einstein 1921, 212; Paty 1992, 7-8), which Einstein takes to mean that only the conjunction of a mathematical structure G with a physical part P is tested experimentally. This implies in particular that concepts like length and duration would not have a direct operational meaning (Dieks 2010, 229). However, Einstein also recognizes that in the present stage of development of physics, a physical theory does not conform to Poincaré's conventionalism (Paty 1993, 304-5). In the context of his philosophy of geometry, Einstein concludes that this situation implies in particular the above-mentioned necessity of taking rods and clocks as independent conceptual elements of

At the present stage, the clock as a (self-sufficient or independent) concept is linked in a very direct way to the clock as an actual measuring instrument, i.e. one compares/attributes to the clock as concept the measurements made by the measuring instrument that as such must have been calibrated to the adopted metrological unit, which was given, in Einstein's days, by the sidereal time (i.e. the rotation of the Earth). Even if in several places Einstein mentions balance wheel clocks (see, e.g., Einstein 1911, 344), which can only be taken as a measuring instrument after being calibrated to the unit of time adopted, there is another type of 'clock' that Einstein mentions (see, e.g., Einstein 1907a, 232; Einstein 1907b, 263; Einstein 1910, 134). Einstein noticed that atoms could be used as clocks. In fact, in my view, Einstein anticipated the idea of atomic clock in his writings. The atoms emit and absorb radiation at particular frequencies. In this way to the atoms of different elements corresponds a particular 'signature' of spectral lines (atomic spectra). According to Einstein:

Since the oscillatory phenomena that produces a spectral line must be viewed as intra-atomic phenomena whose frequencies are uniquely determined by the nature of the ions, we can use these ions as clocks. (Einstein 1910, 124-5)

The crucial aspect of this clock is that (at least in inertial motion) it has always the same frequency(ies). In this way we are free to calibrate it to the sidereal time or use it to define the unit of time, as it was done later with the adoption of the atomic time scale based on a metrological definition of the second in terms of the 'internal oscillations' of the cesium atom.[4] There is another point where Einstein's reference to natural clocks becomes, in my view, quite important. In a couple of places Einstein makes reference to an assumption 'hidden' in the derivation of the Lorentz transformations. When determining the transformations that relates space and time coordinates in two inertial reference frames S and S' in relative motion with velocity $\upsilon$, Einstein still has a function $\varphi(\upsilon)$ undetermined (Einstein 1907b, 260). Einstein considers a third inertial reference frame S" moving with velocity $-\upsilon$ in relation to S'. This means that S and S" coincide. This sets the function $\varphi(\upsilon)$ to 1. At this point Einstein mentions, in a footnote, the following:

---

the theory. The clocks and rods Einstein refers to are above all concepts of the theory, which in the present stage have a very direct link to instrumentation and also to the mathematical structure (this is also the case with Einstein's theory of gravitation; see, e.g., Einstein 1955, 63-4). As conceptual elements, the clocks (and associated notions of duration/time interval/passage, etc.) and rods (and associated notion of length) – and linked as they are with the geometrical structure of the theory – are not used in the context of the theory simply as a representation of particular instruments; they are part of a conceptual-mathematical 'definition' of length and duration, which has a clear counterpart in experimentation via rods and clocks as measuring instruments. As instruments, rods and clocks do not define length and duration, they enable the measurement of length and duration. In conclusion, Einstein's view on rods and clocks as independent conceptual elements of the theory – crucial to the view developed in this work – was developed not in the context of Einstein's early eventual operationalism, but is related to Einstein's later views of geometry as practical/physical geometry. In this way it does not fell prey to any operationalist philosophical position, which as Brown remarks "got such a bad name in philosophy, [that] it has been fashionable for some time in the philosophical literature to discuss space-time structure without any reference at all to such base elements as rods and clocks" (Brown 2005, 95).

[4] When in inertial motion the atomic clocks gives/reads a time that is equal to the inertial time. We can say that experimentally the atomic time scale and the inertial time scale are identical (Ohanian 1976, 187-8).

This conclusion is based on the physical assumption that the length of a measuring rod or the rate of a clock do not undergo any permanent changes if these objects are set in motion and then brought to rest again. (Einstein 1907b, 260)[5, 6]

At this point of the deduction of the Lorentz transformations, Einstein is thinking in terms of an actual setting into relative motion of identically constructed reference frames. Instead of thinking in terms of initially given reference frames that are taken to be in relative motion, Einstein is considering the case, e.g., of setting into motion, in relation to the 'given' inertial reference frame S, the reference frame S', and later the setting into motion of S" in relation to the already inertial reference frame S'. The problem is in that undefined moment in which S' and S" are no more at rest in relation to an inertial reference frame but not yet with a constant velocity in relation to the inertial reference frame, i.e. that moment in which S' and S" cannot properly be consider as inertial reference frames. It is here that the boostability assumption is, implicitly, taken into account.[7]

One might get the impression that the assumption being beyond the principle of relativity is somewhat ad hoc, even if necessary according to Einstein. That might not be the case. There is a way of seeing how this assumption is inbuilt in the theory given by Einstein himself. The justification for this assumption results from the identification of the conceptual clock with the notion of natural clock arising in experimentation. The clock of the theory of relativity is still a self-sufficient or independent concept, but it represents within the theory natural clocks; and the natural clocks to which Einstein relates his conceptual clock are atoms.[8] In this way, the assumption being made with respect to the conceptual clock inbuilt in the theory can be seen as justified when taking into account experimental results about natural clocks. According to Einstein:

---

[5] Einstein also refers to the assumption as the "independence of measuring rods and clocks from their past history" (Einstein 1920, 127).

[6] Brown refers to this assumption as the boostability of rods and clocks (Brown 2005, 30). Brown is clear in the dynamical aspect of this assumption (Brown 2005, 28). However the rods and clocks as 'primitive' conceptual elements are not dynamically described in the theory. This means that any dynamical considerations being made, e.g. by reference to applied forces (Brown 2005, 95), are not rigorous, having more a heuristic role.

[7] As a dynamical assumption, even if specific to the particular situation of a boosted reference frame, this assumption does not follows from the principle of relativity of inertial motion, according to which "the laws governing natural phenomena are independent of the state of motion of the [inertial reference frame] with respect to which the phenomena are observed" (Einstein 1910, 123). The principle is circumscribed to a statement of the equivalence of inertial frames for the description of phenomena (in more 'modern' terms it states that the laws of the theory are Lorentz covariant; see, e.g. Norton 1993, 796); as such it does not bear on the issue of the boosting of the inertial reference frame itself. In this way, the boosts of rods and clocks as primitive entities inscribed in the notion of inertial reference frame are outside the implications of the principle of relativity.

[8] Torretti associates a different notion of natural clock to Einstein, the light clock (Torretti 1983, 52-4; Ohanian 1976, 192-3). In this context Torretti refers to Einstein's time scale (Torretti 1983, 54). This time scale cannot be seen as independent from the inertial time scale; as Ohanian call the attention to, the time variable that appears in the equations of electrodynamics corresponds to an inertial time (Ohanian 1976, 195). The situation with the atomic time scale is different. It is correct that we cannot completely separate conceptually the atomic time scale provided by an atom from the fact that, as a 'clock' of an inertial reference frame, its state of motion is taken into account. However it is clear that there is an 'internal' physics of the atom leading to the atomic time that, at the present stage of development of theoretical physics, is taken to be independent from the inertial time. The fact that experimentally the two time scales coincide (Ohanian 1976, 187-8) does not mean that conceptually and in experimentation they are not distinguished.

If two ideal clocks are going at the same rate at any time and at any place (being then in immediate proximity to each other), they will always go at the same rate, no matter where and when they are again compared with each other at one place. If this law was not valid for natural clocks, the proper frequencies for the separate atoms of the same chemical element would not be in such exact agreement as experience demonstrates. (Einstein 1921, 213-4)

These remarks are made in the context of a discussion about the so-called practical geometry. In this way we do not have to read them as referring strictly to the boostability of clocks. In these remarks Einstein might be considering a more general situation than boostability; that of general accelerations of the clocks between two inertial states. However this means that the assumption being made regarding the boostability of clocks (that is very specific case of the more general acceleration of clocks) can be seen as grounded in experimental findings regarding natural clocks.

    In fact the experimental results go beyond boosts and the more general case referred. In this way, we can give to the assumption of the 'independence of measuring rods and clocks from their past history' a meaning more general than Einstein's original one. Experimentally, the rate of natural clocks, at any particular moment, turns out to be independent of their past non-inertial motion even if its actual motion is still non-inertial, i.e. the rate of natural clocks is not affected, in a direct way, by their acceleration (see, e.g., Zhang 1997, 190-4). The boostability assumption finds its justification in the experimental results that the rates of natural clocks are independent of their past history in this generalized sense.

    The almost implicit adoption of this assumption regarding the conceptual clocks of the theory can be seen as the implementation within the structure of the theory of an experimental finding regarding natural clocks that is previous (or at least independent) to the eventual theoretical 'allocation' of these experimental results using the theory of relativity.

    In my view, with the notion of natural clock Einstein is anticipating the notion of atomic time that enabled the adoption of an atomic time scale. A natural clock has an empirical (proper) time that stands on its own, not having to be calibrated to any motion in the 'heavens'. In fact it turns out that atomic time enables a more accurate determination of the second (metrologically defined as the elapsed time of 9192631770 oscillations of the cesium atom) than the Earth's rotation or the so-called ephemeris time (Jespersen and Fitz-Randolph 1999, 110). I will try to show in this paper the importance of this notion to clarify several conceptual issues related in particular to proper time, and the so-called clock hypothesis.

2 Brown's views on proper time and the clock hypothesis

Brown applies the term proper time to mean the time read off by a real or conceptual clock (see, e.g. Brown 2005, 29, 115). As we have seen the conceptual clock is the counterpart within the theory of the natural (atomic) clocks. This means that ultimately, in Brown's account, proper time refers to the empirical time given by a real clock. Brown refers to the conceptual clock, initially for the case of inertial motion, as the ideal clock, and so refers to the proper time of ideal clocks. The conceptual clocks are ideal in the sense of being clocks (in inertial motion) for which a free particle takes equal time intervals to traverse equal distances. In this way ideal clocks are clocks that give a time reading according to the law of inertia (Brown 2005, 19), i.e. ideal clocks are clocks in inertial motion that give the universal inertial time (Brown 2005, 95).[9]

Brown follows Einstein in his definition of coordinate time. One takes clocks at rest in an inertial reference frame, properly synchronized; the time coordinate is the time read by these stationary clocks (Brown 2005, 7, 19).

It is important to notice that with Minkowski, proper time is the name given to the interval along a time-like worldline, and it is taken for granted that the duration read by clock is given by the Minkowski proper time.[10] With Brown, proper time is simply the time (number of 'cycles') read off by a clock, which might or might not be equal to the interval along the clock's worldline (Brown 2005, 95, 115): proper time is the duration read off by a clock independently of its state of motion being inertial or accelerated. According to Brown, if the clock hypothesis[11] applies to a particular clock (with a particular acceleration), "then the clock's proper time will be proportional to the Minkowski distance along its world-line" (Brown 2005, 95).

This view is inconsistent with another of Brown's claims. As we have seen, the coordinate time is defined by synchronizing identical clocks sharing the same inertial motion. Brown maintains (in part implicitly) that there is a difference between proper time (in Brown's sense) and coordinate time (see, e.g., Brown 2005, 7, 19, 29), with which I agree. However Brown claims that coordinate time is more fundamental than proper time (Brown 2005, 92). That cannot be the case. According to Brown's own 'definition', the time being read by clocks, in inertial motion or not, is called proper time. The coordinate time results from synchronizing clocks, in inertial motion, with an identical proper time, i.e. the coordinate time is built in terms of the proper time (in Brown's sense) of the clocks of the inertial reference frame. It cannot be the case that the coordinate time is more fundamental than proper time as defined in this way.

In Brown's view, the identification of the interval of a time-like worldline with the time being measured/given by a clock on this worldline needs for its justification the clock hypothesis. This is a dynamical condition that a clock must follow (more exactly, its theoretical model). Brown makes his point in terms of a classical picture based on the concept of force: a clock is seen as a complex dynamical system whose cyclic behaviour is described by taking it to be an isolated system described dynamically by internal forces. If the external forces that accelerate the clock along a particular time-like worldline are small in comparison to the internal forces they will not affect the cyclic behaviour of the clock and because of this its timekeeping. Under this condition the time being given by the clock's cycles (that Brown calls the proper time) is identical to the interval along the worldline (Brown 2005, 95).

---

[9] In a couple of places Brown uses the term inertial clock instead of ideal clock to refer to a (conceptual) clock in inertial motion giving the inertial time (Brown 2005, 19, 97). To simplify the terminology adopted in this work, and avoid confusion with Arthur's terminology, I will simply refer to conceptual clocks.

[10] As it is well known, Minkowski presented the notion of proper time in 1908. By definition the proper time is associated just to substantial material systems to which one can associate time-like worldlines. For this particular case one considers the invariant infinitesimal interval along the worldline at the position of the material system $c^2 d\tau^2 = c^2 dt^2 - dx^2 - dy^2 - dz^2$. According to Minkowski, "The integral $\int d\tau = \tau$ of this amount, taken along the world-line from any fixed starting-point $P_0$ to the variable endpoint P, we call the "proper-time" of the substantial point in P" (Minkowski 1908, 85). Immediately after defining proper time, Minkowski determines the motion-vector and acceleration-vector using $d\tau$ as the infinitesimal time interval gone by the material system.

[11] In Brown's own words, the clock hypothesis is "the claim that when a clock is accelerated, the effect of motion on the rate of the clock is no more than that associated with its instantaneous velocity – the acceleration adds nothing" (Brown 2005, 9). According to Brown this is the case when "the external forces accelerating the clock are small in relation to the internal 'restoring' forces at work inside the clock" (Brown 2005, 115).

By applying the clock hypothesis to the conceptual clock as a physical (dynamical) system, one has to show: (1) that acceleration has no effect on the rate of the clock, i.e. the effect of motion on the rate of the clock is no more than that associated with the instantaneous velocity (Brown 2005, 9); (2) that in this situation it follows straightforwardly from the theory that the proper time of the clock is identical to the Minkowski interval along the time-like worldline (Brown 2005, 95). In this way, it is supposed to follow directly that clocks undisturbed by an acceleration give a time reading according to Minkowski's definition of proper time, as the integral $\tau = \int ds/c$ along a time-like worldline, where $ds^2 = c^2 dt^2 - dx^2 - dy^2 - dz^2$. To simplify let us consider a clock accelerated along the x axis; we can write ds as $dt \sqrt{1 - (dx/dt)^2/c^2}$, i.e. $ds = dt \sqrt{1 - \upsilon^2/c^2}$, where $\upsilon$ is the instantaneous velocity of the accelerated clock (see, e.g. Bohm 1965, 163). In this way we have, for the integral along a time-like worldline that Minkowski called proper time, $\tau = \int \sqrt{1 - \upsilon^2/c^2}\, dt$. Importantly, $\int ds/c$ was calculated without any reference to the time reading of the accelerated clock.

To establish a connection between $\int ds/c$ and the time reading of the accelerated clock, one considers an infinite set/sequence of inertial reference frames, each one located at a point of the clock's trajectory and with a velocity equal to the clock's instantaneous velocity at that point (see, e.g. Bohm 1965, 162-3). For each one of these inertial reference frames, let dt' be the time reading of a clock of the inertial reference frame located momentarily side by side with the accelerated clock; in this case ds' = c dt'. According to the theory of relativity ds' = ds, i.e. $dt' = dt \sqrt{1 - \upsilon^2/c^2}$. In this way we establish a relation between the time readings of clocks of two inertial reference frames (i.e. a relation between coordinate times), one in which the worldline of the accelerated clock is described an another in which the accelerated clock is momentarily at rest.

It is here that the connexion with the time reading of the accelerated clock is made. One takes the rate of the accelerated clock not to be affected by the acceleration, i.e. by the non-inertial motion. In this way, the accelerated clock and the momentarily co-moving clock (in inertial motion) have the same rate. This means that we identify, for each momentarily co-moving inertial reference frame, dt' with the time measured by the accelerated clock. Since $dt' = dt \sqrt{1 - \upsilon^2/c^2}$, the rate of the accelerated clock only depends on its instantaneous velocity (as measured in the inertial reference frame in which the worldline of the accelerated clock is described). In Brown's view, this connexion of the time reading of the accelerated clock with the time reading of a momentarily co-moving clock in inertial motion is made by resort to a dynamical assumption regarding the workings of the accelerated clock, which is not a consequence of the postulates of the theory of relativity, the clock hypothesis (see, e.g., Brown and Pooley 2001, 264-5).

The final step is to 'sum' over the time readings of all the momentarily co-moving (inertial) clocks, arriving at the same mathematical expression (an integral) that Minkowski called the proper time. In this 'derivation' we are suppose to be showing the identity between an integral $\int ds/c$ calculated in terms of the coordinates of a reference frame and an infinite summation of infinitesimal elements ds'/c (i.e. dt') of different inertial reference frames, corresponding to time readings of different clocks belonging each to a different frame. Following Brown we might say that the clock hypothesis enables to identify for each frame the time reading of the accelerated clock with these elements dt'.[12] In this way, the total time reading of the accelerated clock is equal to the summation $dt_1' + dt_2' + dt_3' + \ldots$. For a large number of co-moving inertial reference frames, this sum is supposed to approach the Minkowski integral. What we have then is an infinite summation $\sum dt_n' = \sum \sqrt{1 - v_n^2/c^2}\,dt$, where $v_n$ is the instantaneous velocity of the accelerated clock when momentarily at rest in relation to the inertial reference frame number n. We take this summation to be replaceable by the integral $\int \sqrt{1-v^2/c^2}\,dt$, which corresponds to the Minkowski definition of proper time.

As this 'derivation' shows, it seems necessary a dynamical assumption to justify attributing to an accelerated clock the same rate as a clock in inertial motion in relation to which it is momentarily at rest. This is, in Brown's view, an extra condition that a clock must satisfy in order to identify its time reading (called by Brown proper time) with the integral of the invariant infinitesimal interval along a time-like worldline (called by Minkowski proper time).

3 Arthur's views on proper time and the clock hypothesis

To Arthur, Minkowski most enduring contribution to physics was, more than the definition, the discovery that the theory of relativity enables a new notion of time that Minkowski called the proper time (Arthur 2010, 159). In Arthur's view the theory of relativity brings with it a degeneracy of the concept of time in two different concepts: proper time and coordinate time (Arthur 2008, 207-8). The importance of this bifurcation can be seen in Arthur's view when addressing the issue of what notion of becoming the theory enables. According to Arthur:

in [the theory of relativity] the becoming of events in succession, the rate of a process or the rate at which a thing ages, is tracked by proper time; the synchronization of distant events is tracked by the time-coordinate function (Arthur 2008, 208).[13, 14]

It is the time elapsing along a particular path in spacetime [(i.e. the proper time)] that measures how fast the processes traversing that path are going, how fast the people or things undergoing them are ageing, how fast they are becoming. (Arthur 2008, 217)

To Arthur, proper time is a physical quantity predicted by the theory of relativity, since it is invariant, i.e. independent of the particular inertial reference frame adopted (Arthur 2010, 177). This temporal physical quantity associated with a (material) physical system

---

[12] This sentence is applicable both in Brown's as in Arthur's case; both take into account the clock hypothesis even if disagree on its 'place' in the theory. My position is a bit different. For the case of natural clocks, their behaviour under acceleration is an input physical assumption that we ascribe to their conceptual counterpart within the theory, i.e. we do not need any reference to a clock hypothesis in the case of natural clocks. Its 'utility' would be restricted only to the case, e.g., of mechanical clocks like the balance wheel clock or the pendulum.

[13] The time coordinate at a particular location in a real/conceptual inertial reference frame can be seen as given by the time reading of a natural/conceptual clock (that we take to have been synchronized with the rest of natural/conceptual clocks of the inertial reference frame). Since an event, like a thunder striking next to the clock, as a space-time point is given by the time reading of the clock and its position in the inertial reference frame, when we talk about the synchronization of distant events we are actually referring to the synchronization of the natural/conceptual clocks in inertial motion. In this way, it is the shared inertial/atomic time of all the clocks of the inertial reference frame that enables the definition of the time coordinate (by synchronizing the clocks), and from this the synchronization of distant events. In this way the bifurcation of time in the theory would ultimately rely on the distinction not of coordinate time and the Minkowski proper time, but of the atomic/inertial time from the Minkowski proper time.

[14] Several authors defend the view that becoming in the theory of relativity is tracked by the proper time (see, e.g., Dieks 1988, 2006; Savitt 2011)

is calculated as an integral along the time-like worldline of the physical system (i.e. as the Minkowski proper time). It is important to stress the term 'calculated' instead of 'measured'. In Arthur's view proper time is defined in the theory and as such it is extracted from its mathematical structure by performing the above-mentioned calculation along a time-like worldline.

To Arthur it is clear that "as defined, the proper time cannot be evaluated without adopting a system of coordinates" (Arthur 2010, 161). However in his view this does not leads to any conceptual dependence of proper time in relation to the inertial (coordinate) time. According to Arthur, for the time elapsed along any worldline [the proper time] gives a measure that is independent of the co-ordinates, even if a particular frame must be adopted in order to calculate its value. (Arthur 2010, 161)[15]

To Arthur there is a notion that is predicted by the theory; that of ideal clock (Arthur 2010, 159):[16] an ideal clock is one that reads proper time (Arthur 2010, 166). However Arthur also writes that "in the context of the global Minkowski spacetime of [the theory of relativity] the fact that an ideal clock indicates proper time follows straightforwardly" (Arthur 2010, 172). I think that in this sentence when Arthur refers to 'ideal clock' he is actually thinking about theoretical models of a real clock that behaves as an ideal clock. This is the only way I can make sense of this sentence without being in contradiction with Arthur's view that ideal clocks reading Minkowski's proper time are predicted or implicitly determined by the theory. In fact this reading of Arthur also enables to make sense of his remarks about the clock hypothesis. According to Arthur:

[The clock hypothesis] does not have the status of an independent hypothesis, but is simply a description of the behaviour of an ideal clock as predicted by [the theory of relativity]". (Arthur 2010, 159)

The [clock hypothesis] is not needed as an independent postulate in [the theory of relativity]. Insofar as it can be regarded as stating the criterion for an ideal clock in [the theory of relativity], it is already implicit in that theory in the invariance of proper time. (Arthur 2010, 177)

Thinking in terms of a more or less heuristic model of a clock whose time reading when accelerated his equal to the invariant Minkowski interval, the clock hypothesis is a very general criterion – stated in terms of a very general description of the dynamical behaviour of clocks –, that any particular accelerated clock most follow to behave as an ideal clock (implicitly determined by the invariant Minkowski interval). This is the case when the effect of the acceleration on the workings of a clock can be considered as very small in relation to the 'internal forces' at work within the clock. One example of an ideal clock is, according to Arthur, an atomic clock (Arthur 2010, 168).

---

[15] Arthur even contends that "the whole content of relativity theory can now be framed in terms of [the proper time], so that co-ordinates are no longer regarded as primitive" (Arthur 2010, 161-2). It is important to notice that only along the time-like worldline can the coordinates be seen as functions of proper time, when taken to be parameterised by the proper time. It is only in this strict sense that one should read Arthur's reference to the coordinates as no longer primitive. Minkowski's presentation is clearer on this. According to him, "*on the world-line* we regard x, y, z, t ... as functions of the proper time τ" (Minkowski 1908, 85 [my emphasis]).

[16] Elsewhere Arthur writes that an ideal clock is "implicitly determined by theory" (Arthur 2010, 168).

Arthur is clear in separating the clock hypothesis – taken to be very general dynamical criterion that he takes to be implicit in the theory – from the issue of whether a particular physical system will behave as an ideal clock (Arthur 2010, 167-8). According to Arthur "the argument that many real clocks will fail to satisfy the clock hypothesis is just the claim that many processes fail to qualify as ideal clocks" (Arthur 2010, 177).

However this brings an issue that Arthur does not address. According to Arthur the becoming of physical systems is given/tracked by the Minkowski proper time. Since it is quite easy to have physical systems that when accelerated will not behave as ideal clocks, one example being the pendulum clock (see, e.g., Arthur 2010, 167), we face the problem of how to track the becoming for these physical systems that do not behave as ideal clocks? Is it still given by the Minkowski proper time or, e.g., by the 'cycles' of the physical system?

4 A different view in terms of the notion of natural clock

As we have seen, the conceptual clock as a concept of the theory of relativity has implicit boostability as a dynamical assumption. This is not simply a necessary but ad hoc theoretical assumption. It can be seen as arising from an experimental finding, that the rate of natural clocks (e.g. atoms) does not depend on their past history (see, e.g., Zhang 1997, 190-4). This justifies ascribing the boostability to conceptual clocks, as the counterpart within the theory of natural clocks.

Taking explicitly into account the completed theory of relativity, it turns out that the independence of past history means that natural clocks read proper time as defined by Minkowski. Taking the conceptual clock as the conceptual counterpart of the natural clock, it must also be an ideal clock in Arthur's sense of having a time reading whose value is identical to the Minkowski proper time when in non-inertial motion.

As we have seen, the coordinate time is defined by synchronizing natural/conceptual clocks of an inertial reference frame, i.e. by setting the phase of clocks sharing a universal atomic/inertial time. Following Arthur we might say that the construction of the theory of relativity led to the appearance of another temporal physical quantity that is built in terms of the system of coordinates but that is independent of any particular system of coordinates chosen: the Minkowski proper time. This would warrant, according to Arthur, the bifurcation of time.

The problem with this view is that the theoretical construction of the theory is made by taking into account clocks to which the boostability assumption applies, i.e. by taking into account experimental results regarding the empirical proper time of atomic clocks. It is correct that we do not use as a physical input assumption the experimental result that the empirical proper time of atomic clocks under acceleration corresponds to the Minkowski proper time. Before the theory was built we did not have any notion of Minkowski proper time, and historically the more general experiments related to accelerated physical systems that I am making reference to where made after the development of the theory of relativity.[17] However it is difficult to maintain a complete independence of the Minkowski proper time as a physical quantity from the experimental results about the proper time of atomic clocks since these are in part taken into account in the development of the theory; specifically, according to Einstein, in the notions of inertial reference frame and in the deduction of the Lorentz transformations (Einstein 1907b, 260, 263; Einstein 1910, 134).

---

[17] As we have seen, Einstein only mentions the 'independence from past history' in the more specific sense that identical clocks that are again in relative inertial rest have the same rate independently of their past motions.

In my view, the coordinate time and in part the Minkowski proper time, can be seen as arising from taking into account in the theory's construction the empirical (proper) time of natural clocks. In this way, there is another relevant notion of time at work in the theory of relativity that of empirical proper time of natural clocks.

In fact if we accept a clear-cut bifurcation of time in the theory of relativity in terms of the coordinate (inertial) time and the Minkowski proper time we soon face difficulties. Let us for the sake of the argument accept Arthur's view on becoming. As we have seen the bifurcation of time according to Arthur leads to very different roles to the coordinate (inertial) time and the Minkowski proper time; the becoming of physical systems is given/tracked by the Minkowski proper time, and the synchronization of distant events is given by the time coordinate (Arthur 2008, 208).

Let us consider a clock in inertial motion. According to Arthur, the becoming of the clock is tracked by/calculated using the Minkowski proper time. However, in this case the Minkowski proper time has the same value as the time coordinate of the clock under consideration (i.e. the clock's atomic/inertial time). To Arthur, "this is only numerical equality, not identity" (Arthur 2008, 219).

From the perspective of a clock as a natural (atomic) clock, its time reading (i.e. its empirical proper time as given by the clock's 'cycles') is ascribed to the clock as a conceptual element of the theory. In the case of an inertial motion, following Arthur, this would give rise only to the coordinate (inertial) time not the 'number of the becoming' of the atomic/conceptual clock; the becoming would be given/tracked by the Minkowski proper time that in the case of an inertial motion is numerically identical to the time coordinate of the clock. In terms of the atomic clock, the Minkowski proper time would be identical to the empirical proper time, but in Arthur's view this identity would not point to the possibility that the numerical description of the becoming of the atomic clock in the theory has its 'root' in the empirical proper time of the atomic clock that corresponds to the atomic/inertial time (in the case of an inertial motion); the atomic clock's becoming would be derived/calculated via theory through the invariant Minkowski interval. In my view this is not correct. To see why let us begin by considering the case of accelerated clocks. I will focus on two types of clocks: natural clocks, which in the case of the theory of relativity means atomic clocks (more exactly, atoms), and mechanical clocks like the pendulum.

The atomic clock and the pendulum behave in a very different way under 'strong' accelerations. Experimentally, the atomic clock's empirical proper time is identical to the Minkowski proper time (see, e.g., Zhang 1997, 190-4). In the case of the pendulum, for moderate accelerations it gives a time reading according to the Minkowski proper time but under a stronger acceleration the empirical proper time of the pendulum is different from the Minkowski proper time (Arthur 2010, 167). Let us consider that there is an argument to maintain that even in this case we might consider that the becoming of the pendulum is tracked by the Minkowski proper time (Arthur does not present this argument but I will give it here). This would seem as confirming Arthur's view, since the becoming would be given by the Minkowski proper time independently of the time reading of a clock. In the case of inertial motions the time reading of a clock is identical to the Minkowski proper time as in the case of what Arthur called ideal clocks. But there are also physical systems that do not behave as ideal clocks, which nonetheless have their becoming given by the Minkowski proper time. This would seem to vindicate Arthur's views: the empirical proper time of an accelerated clock might be different from the Minkowski proper time that gives the actual time gone by the clock (in Arthur's view tracking the becoming of the clock). Only in the case of an inertial motion there is an apparent conflation of the Minkowski proper time with the coordinate (inertial) time corresponding to the empirical proper time.

However there is a loophole in this argumentation. I treated atomic clocks and physical systems, like the pendulum, at an equal footing. But this cannot be done. As Einstein mentioned, the clocks associated to the notion of time (in the inertial reference frames), enter the theory as self-sufficient and theoretically independent conceptual elements of the theory. The atomic clock as a 'source' of atomic time (in inertial motion) is not theoretically describe in the theory of relativity; on the contrary, as we have seen, there are physical input assumptions being made (and ascribed to clocks as conceptual elements of the theory), by taking into account experimental results about atomic clocks. To consider that the Minkowski proper time gives the becoming of atomic clocks is to reverse the order of things.[18] We have atomic clocks as fundamental elements of the theory. We must consider, when adopting Arthur's views, that their empirical proper time is marking the becoming of the atomic clocks. This implies that the time coordinate of the clocks in inertial motion already gives the numerical value of the becoming. In the case of an accelerated motion, the proper time of accelerated atomic clock is different from the shared proper time of all atomic clocks in inertial motion; there is a 'slowing down' of the accelerated atomic clock's becoming in relation to the becoming of the atomic clocks in inertial motion (see, e.g., Arthur 2010, 164). The theory provides an invariant expression that enables to calculate the atomic clock's empirical proper time. This is the Minkowski proper time. This is quite fantastic since, as mentioned, we did not take into account all the experimental knowledge about accelerated atomic clocks to formulate the theory. Regarding acceleration the only assumption inbuilt in the theory was the boostability assumption. However the fact that the theory has the Minkowski proper time as an invariant quantity does not make this quantity more fundamental than the empirical proper time of atomic clocks that we started with in the first place.

Let us now return to the issue of physical systems, like the pendulum, that when accelerated do not behave as ideal clocks. The 'root' of time in the theory of relativity is in the natural clocks, i.e. atoms. The pendulum or any other complex system can be seen as constituted by these simpler building blocks that within the theory of relativity are conceptualised in classical terms and have associated a trajectory in space and time. These 'classical atoms', i.e. the conceptual clocks, are not affected by the acceleration, they still have a time reading equal to the Minkowski proper time. There are situations in which the pendulum does not behave any more as an ideal clock in Arthur's sense (i.e. its time keeping does not correspond to the numerical value of the Minkowski proper time). In this case it might look as if its worldline looses a metrological interpretation in terms of the Minkowski proper time. However if we, so to speak, magnify the worldline of the complex system, and represent the worldlines of the constituting 'classical atoms', these enable to maintain the meaning of the Minkowski interval for each of these 'fundamental' worldlines. In this situation the Minkowski proper time of the worldline of the complex clock gives the Minkowski proper time of the constituting 'classical atoms'; in Arthur's view, it tracks the becoming of the

---

[18] As mentioned in section 1, the atomic/natural clock and related notion of time (metrologically implemented through the atomic time scale) is independent of the formulation of the theory and in fact is taken into account in input physical assumptions of the theory. It is from measurements made with accelerated atomic clocks that we check that the Minkowski proper time in fact gives the time gone by an atomic system. Not the other way around.

'classical atoms', and because of this also the becoming of the complex system made of 'classical atoms'.[19]

From this we can conclude two things: (1) not taking into account some quantum phenomena that is outside the scope of the theory, in all cases the Minkowski proper time can be taken to give the time gone by a physical system (in Arthur's terms marking the becoming of the physical systems); (2) Since this happens even for physical systems whose time reading does not correspond to the Minkowski proper time, as a tool to determine the actual time gone by a physical system, the Minkowski proper time (i.e. the invariant integral calculated along the time-like worldline of a physical system) enables to calculate the time gone by a physical systems 'on top of' the actual time reading we might have with the physical system.

In my view it is this incredible applicability of the Minkowski proper time that we extend even to physical systems that do not provided any measure of time (i.e. physical systems that cannot be used as clocks) – since we are considering physical systems made of atoms – that gives rise to the view of the Minkowski proper time as an almost fundamental physical quantity predicted by the theory of relativity, giving the time gone by physical systems for inertial and non-inertial motions. The only thing wrong in this view would be in the adjective 'fundamental', since 'behind' and 'sustaining' the Minkowski proper time is the empirical proper time of atoms.

This view has implications regarding the clock hypothesis. As the counterpart of the atomic clock we need for a question of consistency that the time reading that we attribute in the theory to the conceptual clock in non-inertial motion (i.e. its proper time) to be equal to the empirical proper time of an accelerated atomic clock. It turns out that the theory of relativity provides through the invariance of the Minkowski interval a theoretical slot where to fit the experimental result of natural clocks whose rate does not depend on their past history (i.e. it does not depend in a direct way on its acceleration). In this way, the ideal clock in Arthur's sense already emerges as the conceptual clock as a self-sufficient theoretically independent concept taken to be the counterpart, within the theory, of natural clocks: the time reading (proper time) of the atomic/conceptual clock is calculate with/given by the Minkowski proper time.

It is true that we might wonder why the natural clock is like that, and try to develop a theoretical model that explains why a natural clock does not depend on its past history. However the notion of natural clock is an input assumption of the theory, and in fact its 'dynamical' description, explaining, eventually, the independence of its rate on the acceleration, must be made totally by reference to theories that are beyond the scope of the theory of relativity (at the moment quantum theories). This is not the clock hypothesis. In fact the clock hypothesis is a very general dynamical rule of thumb, applied within the theory of relativity to the models of real clocks (sometimes making, in part, a heuristic reference to other theories like quantum mechanics); as such it does not address natural clocks.[20, 21]

---

[19] There are however simple physical systems that cannot be seen as constituted by simpler 'classical atoms', which in certain cases do not behave as ideal clocks. One example of a theoretical model with this characteristic is a type of flavour oscillation clock that when subjected to a rotational acceleration does not read proper time (Knox 2010). However, in this case, we are considering a theoretical model (with an eventual real counterpart) that is outside the scope of the theory of relativity. Even the idea of worldline cannot be applied in the case of a quantum system like the flavour oscillation clock. The natural clock (not depending on its past history) is conceptualised as an ideal classical clock with a classical worldline. Quantum systems that do not behave as ideal clocks open interesting questions regarding the notion of becoming as something that can be tracked by the Minkowski proper time; however this is outside the scope of the theory of relativity and the views being presented/analysed in this work are only related to this theory.

In this way the relevance of the clock hypothesis is much smaller in my account than in Arthur's. We might want to apply it case-by-case to models of real clocks, but it does not have the centrality that it has in Arthur's account. We do not have to take it as implicit in the theory because of the invariant Minkowski interval. This is already 'covered' by the conceptual clock as the counterpart of the natural clock, which is an ideal clock in Arthur's sense. Also, when adopting Arthur's interpretation of the Minkowski proper time as giving the becoming of physical systems, the 'becoming' of physical systems is still given by the Minkowski proper time even if these do not behave as ideal clocks. The 'clock hypothesis' would simply provide a general rule of thumb to predict when the functioning of a physical system as a clock (i.e. its 'cycles') will correspond to its actual becoming (as given by the Minkowski proper time).

In the same way, I would not take the clock hypothesis to be a sort of 'extra assumption' as Brown's view might lead us to think. As mentioned, it turns out to be a sort of very general 'dynamical recipe' for a direct heuristic application of a simple dynamical reasoning within the theory for the case of real clocks like the balance wheel clock (i.e. clocks that are not taken into account in the development of the theory as the atomic clock), to determine when one can expect these clocks to behave like natural clocks (i.e. to have a time reading when accelerated in accordance to the Minkowski proper time). As a dynamical recipe or rule of thumb, the clock hypothesis is too general and follows too directly from the theory's dynamics to consider it as a dynamical assumption that "is not a consequence of the postulates of the theory of relativity" (see, e.g., Brown and Pooley 2001, 264-5).

Also, like in Arthur's case, the relevance of the clock hypothesis is much smaller in my account than in Brown's. We do not need it, like in Brown's case, to show that we can take the proper time of conceptual clocks as being given by the Minkowski invariant interval. The conceptual clock that corresponds to the natural clock within the theory, as such, is already an ideal clock reading the Minkowski proper time. Also by considering physical systems as constituted by atoms (conceptualized in classical terms within the theory), the time gone by the physical system is given by the Minkowski proper time even if when accelerated the physical system as a clock might be malfunctioning (i.e. its cycles not being identical to the value of the Minkowski proper time). Its 'utility' is simply to predict up to what acceleration we can expect a real clock to give a time reading identical to the actual time gone by the clock (that is calculated using the Minkowski interval).

---

[20] One should not be mislead by references to atomic clocks in the context of discussions regarding the clock hypothesis (see, e.g., Brown 2005, 94-5, Arthur 2010, 168). For example, in the case of the pendulum one is applying basic dynamics of the theory of relativity (see, e.g., Misner et al. 1973, 394-5). This corresponds to the clock hypothesis as a general (dynamical) rule of thumb expressed in terms of the concept of force (or potential). In the case of the atomic clock we find physical 'toy models' in which the clock hypothesis is 'applied' in a completely heuristic way (see, e.g., Ohanian 1976, 207-8). As mentioned, taken the 'clock' as a self-sufficient concept to be the counterpart within the theory of the atomic clock that experimentally we find to give a time reading (i.e. to have an empirical proper time) according to the Minkowski proper time, this fact becomes a physical input assumption of the theory that is not explained within the theory of relativity.

[21] It is my view that the quantum mechanical description of atoms does not explain the independence on its past history of natural clocks. Quantum mechanics is 'built' on top of the notion of inertial reference frame with its implicit conceptual rods and clocks. As Dickson mentions, "just as classical physics does, quantum physics contains an assumption (usually left implicit) that there is some frame (some system of coordinates) in which the laws are valid" (Dickson 2004, 201). However it is beyond the scope of this work to develop this issue here. Even if that was not the case it does not affect the views being defended here in relation to the clock hypothesis.

## 5 Further comments

In this work, by taking into account Einstein's notion of natural clock and the related boostability assumption, I presented a view on the so-called clock hypothesis, different from Brown's and Arthur's. My treatment of the clock hypothesis goes hand in hand with a re-evaluation of the notion of proper time in terms of the empirical proper time of real clocks, something that is already present in Brown's work. The difference with Brown is that the empirical proper time is 'thought' in terms of the time being given by natural clocks as defined by Einstein: physical systems whose 'internal changes' are independent of their past motion. Natural clocks enter the theory's structure in the articulation of the time coordinate of an inertial reference frame and almost as a hidden assumption – the boostability assumption –, necessary in Einstein's view to derive the Lorentz transformations. As such, the existence of natural clocks is not something that is predicted by the theory but a physical input assumption.

In my view, regarding this point, the great merit of Einstein's theory of relativity is making of its weakness its strength. As we have seen, Einstein was unhappy with the fact that the theory relies on rods and clocks as self-sufficient conceptual elements necessary to implement the notion of inertial reference frame. However as Minkowski's contribution made explicit, the theory is able to 'find a place' in its theoretical structure to a concept of clock that corresponds to the experimental notion of natural clocks. With the Minkowski proper time we can fit into the theory the empirical proper time of natural clocks that is (at least in part) implicit in its construction.